# Magnetization dynamics modulated by Dzyaloshinskii-Moriya interaction in the double-interface spin transfer torque magnetic tunnel junction

Simin Li[1], Zhaohao Wang[1,2,3*], Yijie Wang[4], Mengxing Wang[5], Weisheng Zhao[1,2,3*]

[1]School of Microelectronics, Fert Beijing Research Institute, School of Electronics and Information Engineering, Beihang University, Beijing 100191, China

[2]Beijing Advanced Innovation Center for Big Data and Brain Computing (BDBC), Beihang University, Beijing 100191, China

[3]Beihang-Geortek Joint Microelectronics Institute, Qingdao Research Institute, Beihang University, Qingdao 266100, China

[4]Beijing Institute, Beihang University, Beijing 100191, China

[5]School of Instrumentation and Optoelectronic Engineering, Beihang University, Beijing 100191, China

*Correspondence: zhaohao.wang@buaa.edu.cn; weisheng.zhao@buaa.edu.cn

**Abstract**

Currently double-interface magnetic tunnel junctions (MTJs) have been developed for enhancing the thermal stability barrier in small technology node. Dzyaloshinskii-Moriya interaction (DMI) inevitably exists in such devices due to the use of the heavy-metal/ferromagnet structures. Previous studies have demonstrated the detrimental effect of DMI on the conventional single-interface spin transfer torque (STT) MTJs. Here in this work we will prove that the detrimental effect of the DMI could be almost eliminated in the double-interface STT-MTJ. This conclusion is attributed to the suppressing effect of the Ruderman–Kittel–Kasuya–Yosida (RKKY) interaction on the DMI. Detailed mechanisms are analyzed based on the theoretical models and micromagnetic simulation results. Our work highlights the importance of appropriately controlling the DMI in two free layers of the double-interface STT-MTJ.

**Introduction**

Magnetic random-access memory (MRAM) is one of the most promising candidates for the next generation non-volatile memory thanks to its low power consumption, high density, fast access speed, almost infinite endurance, and good compatibility with CMOS technology [1-2]. The elementary device of the MRAM is the magnetic tunnel junction (MTJ), which is composed of a tunnel barrier sandwiched between two ferromagnetic layers (named pinned layer and free layer). Benefiting from the progresses in the perpendicular anisotropy, the feature size of the MTJ has been scaled to below 40 nm or even 1X nm [3-5]. However, a challenge for the sub-40 nm MTJ is to keep the adequate thermal stability barrier $E = \mu_0 M_s H_k V/2$. (with $\mu_0$ the vacuum magnetic permeability, $M_s$ the saturation magnetization, $H_k$ the anisotropy field, $V$ the volume of the free layer). As indicated by this equation, $E$ decreases with the scaling of the MTJ, resulting in a reduction of data retention time. In this context, double-interface MTJs were proposed to achieve high $E$ at the sub-40 nm technology node [6-10]. Using two coupled ferromagnetic layers as the composite free layer, the equivalent volume of $V$ in the double-interface MTJ is increased to improve the thermal stability barrier. Meanwhile, the damping constant is decreased to result in a low switching current.

In double-interface MTJs, ferromagnet/heavy-metal (FM/HM) structure plays an important role in optimizing the performance. On the one hand, FM/HM structure increases the spin-orbit coupling (SOC) to induce the interfacial perpendicular anisotropy. On the other hand, the heavy metal works as a spacer between two ferromagnetic layers of the composite free layer to provide the Ruderman–Kittel–Kasuya–Yosida (RKKY) interaction [11], which ferromagnetically couples the magnetizations of the two ferromagnetic layers in order that they behaves like an identical layer. However, recent works demonstrate that an antisymmetric exchange coupling called Dzyaloshinskii–Moriya interaction (DMI) [12-13] inevitably exists at the FM/HM interface due to the lack of inversion symmetry [14-18]. Therefore the DMI is naturally induced in the double-interface MTJ with FM/HM structures. DMI favors the chiral magnetic textures (e.g. spin spirals, skyrmions, and Neel type domain walls) and dramatically affects the magnetization dynamics, as validated by the recent studies [14-25]. It is important to mention that the role of the DMI will become more complicated in the double-interface MTJ, since two FM/HM interfaces need to be considered together with an additional RKKY interaction. Therefore, it is of significance to reveal the effect of the DMI on the double-interface MTJ.

In this Letter, for the first time, we study the switching process of the double-interface MTJs under the actions of DMI and RKKY interaction. The double-interface MTJ is switched by the spin transfer torque (STT), which is a mainstream approach for the write operation of the MRAM. It was recently reported that the DMI has a detrimental effect on the STT switching [21-22]. Here our results demonstrate that in double-interface MTJs, the detrimental effect of the DMI could be suppressed by RKKY interaction, resulting in a fast switching and more uniform dynamics. Our work proves the robustness of the double-interface STT-MTJ against the negative interfacial effect.

**Micromagnetic model**

The device studied in this work is illustrated in Fig. 1a, with a FM/HM/FM structure as the composite free layer. The HM layer thickness is adjusted to an appropriate value in order that the induced RKKY interaction ferromagnetically couples two FM layers. One of the FM layers is magnetically softer, which is denoted as FL1 (free layer 1), while the other is magnetically harder and denoted as FL2 (free layer 2). To switch the magnetization of the composite free layer, a current is applied to the double-interface MTJ and generates the STT. In this work, we only consider the transmitted STT from reference layer to FL1, whereas the other torques between FL1 and FL2 are neglected. This simplified model is consistent with the previously reported works [26-28]. The DMIs are induced in both FM/HM and HM/FM interfaces and have the opposite signs due to the different chirality [30].

The magnetization dynamics of the FL1 and FL2 in the double-interface MTJ is studied by micromagnetic simulation with OOMMF package. In the codes, we consider the uniaxial perpendicular anisotropy, 6-neighbour exchange energy, DMI field, RKKY interaction, demagnetization field, dipolar interaction, and STT. The parameters and their default values are listed in Table 1, unless stated otherwise.

Typical simulation results of the time-dependent $m_z$ (perpendicular component of the unit magnetization) are shown in Fig. 1b. If the RKKY interaction is sufficiently strong (e.g. σ = $1 \times 10^{-3}$ J/$m^2$ in Case A and Case B), FL1 and FL2 are coupled together and thus their magnetization dynamics are almost identical, no matter whether the DMI is considered or not. It is also seen that the introduction of the DMI distorts the process of the magnetization switching (see Case B), which is in agreement with the reported results [21-23] and can be attributed to the antisymmetric exchange of the DMI. Once the RKKY interaction is not strong enough, the magnetization dynamics of two free layers cannot be ideally coupled so that significant difference between them is observed (see Case C). Below, the simulation results are obtained under a strong enough RKKY interaction, unless stated otherwise.

**Results and Discussion**

Firstly, we study the switching speed under the various RKKY interaction. The switching speed is reflected by a time when $m_z$ reaches 0 (defined as switching time). The $D_1$ and $D_2$ are set to positive and negative values, respectively [30]. The corresponding results are shown in Fig. 2. In the absence of DMI, the switching time increases with the enhanced RKKY interaction, in agreement with the other reported results [26-29]. The reason is that the stronger RKKY interaction makes the magnetization dynamics of FL1 and FL2 more coherently and then equivalently increases the anisotropy. However, the dependence of switching time on the RKKY strength becomes more chaos in the presence of DMI. These results evidence the non-negligible effect of the DMI on the switching behavior of the double-interface MTJ.

Next we study the effect of the DMI in more details. Although $D_1$ and $D_2$ have the opposite signs in reality, in this work we additionally simulate the virtual case when $D_1$ and $D_2$ are of the same sign, for the purpose of comparison. Figure 3 shows the switching time as a function of DMI strength. In Fig. 3a $D_1$ and $D_2$ are set to the same positive values. As can be seen, the switching time rises as the DMI strength increases. This trend is consistent with the reported conclusion that the DMI has a detrimental effect on the STT switching of a standard single-interface MTJ [21-22]. Therefore we infer that the detrimental effects of two positive DMIs are cumulated under the action of ferromagnetically-coupled RKKY interaction. In contrast, such detrimental effects could be mitigated if $D_1$ and $D_2$ are set to the opposite signs, as shown in Fig. 3b, where the variation of switching time is much smaller compared with Fig. 3a. Note that in Fig. 3b the curve is not exactly monotonous, the local fluctuation will be explained later. Remarkably, the effects of DMIs at two interfaces could be cancelled out by appropriately tuning the magnitudes of $D_1$ and $D_2$, as shown in Fig. 3c. These results can be explained in terms of chirality theories as follows.

The DMI energy is expressed as Eq. (1) [31]. As mentioned above, the magnetization dynamics of FL1 and FL2 are almost identical under a sufficiently strong RKKY interaction. In this case, the same $\varepsilon_{DM}$ is obtained in FL1 and FL2. Then the total DMI energy of FL1 and FL2 could be calculated by Eq. (2). Therefore, by setting $D_1/D_2 = -t_2/t_1$, the DMI effects of FL1 and FL2 could be completely offset in the case of a large enough σ, in agreement with Fig. 3c. This conclusion is further verified by the additional results shown in Fig. 3d, where the other parameters are intentionally varied meanwhile keeping $D_1/D_2 = -t_2/t_1$. In addition, it is worth mentioning that Eq. (2) could also explain the results of Fig. 3a, since positive $D_1$ and $D_2$ indeed leads to non-zero DMI which hinders the STT switching.

$$E_{DM} = t \iint D \left[ \left( m_x \frac{\partial m_z}{\partial x} - m_z \frac{\partial m_x}{\partial x} \right) + \left( m_y \frac{\partial m_z}{\partial y} - m_z \frac{\partial m_y}{\partial y} \right) \right] d^2 r = tD\varepsilon_{DM}, \tag{1}$$

$$E_{tot} = (t_1 D_1 + t_2 D_2) \varepsilon_{DM}, \tag{2}$$

$$D_{eq} = \frac{t_1 D_1 + t_2 D_2}{t_1 + t_2}. \tag{3}$$

where $D$ is the continuous DMI constant, $t$ is the thickness of ferromagnetic layer.

The equivalent DMI magnitude ($D_{eq}$) of the composite free layer can be expressed as Eq. (3), which could be used for quantitatively analyzing the effect of DMI on the double-interface MTJ. To validate the effectiveness of Eq. (3), we show the comparison of magnetization dynamics between two pairs of different $\{D_1, D_2\}$ values which lead to the same $D_{eq}$. In Fig. 4a, although there is a little difference between two curves, their overall trends are similar and validate the detrimental effect of the DMI on the STT switching. Here the difference between two curves could be explained as follows. FL1 and FL2 have different anisotropy constants, leading to the local uncertain oscillation of the magnetization dynamics, as shown in Fig. 4c. The same phenomenon is also observed in Fig. 3b. Instead, an ideal case is shown in Fig. 4b and Fig. 4d, where the anisotropy constants of FL1 and FL2 are set to the same values. Clearly, a good coincidence of the two curves is seen, indicating that Eq. (3) could well describe the equivalent DMI magnitude of the double-interface MTJ.

Finally we analyze the time evolution of magnetization dynamics in more details. Figure 5 shows the time-dependent energy during the magnetization switching. The DMI energies of the FL1 and FL2 are accumulated or cancelled, depending on the signs and magnitudes of $D_1$ and $D_2$. This trend is in good agreement with the Eqs. (1)-(3). In addition, the RKKY energies are kept at low values, which validates that the magnetic moments of FL1 and FL2 are synchronously driven. The distributions of RKKY and DMI fields are shown in Fig. 6, where RKKY field plays different roles in various cases. First, the RKKY field in the case of non-zero DMI (see Case 2 and Case 3) is much stronger than that without the DMI (see Case 1). It could be understood that the RKKY field has to overcome the additional non-uniformity of the magnetic textures in the presence of DMI. Second, if $D_1$ and $D_2$ are of opposite signs, the RKKY field resists the DMI fields in both FL1 and FL2 (see Case 2). As a result, the DMI is weakened so that the magnetization dynamics become more uniform. In contrast, once $D_1$ and $D_2$ have the same sign, the RKKY field resists the DMI field in one free layer but assists it in the other free layer (see Case 3). Thus the overall DMI field still has certain effect on the magnetization dynamics, which validates that the DMI cannot be cancelled out if $D_1$ and $D_2$ are of the same sign.

Figure 7 shows the micromagnetic configurations of the FL1 and FL2 during the magnetization switching. Although the domain wall appears in all the cases, different features could be observed at some time moments. It is well known that the DMI favors the non-uniform magnetic textures. Nevertheless, in Fig. 7 uniform magnetization is still formed even in the presence of DMI (see the time when $m_z = -0.5$ in Case 2), as long as the DMI effect is cancelled out. Again, this result validates the theoretical model expressed by Eq. (1)~(3). In addition, it is also seen that the magnetization dynamics is more non-uniform if $D_1$ and $D_2$ are of the same sign (see Case 3 where the domain wall always appears), consistent with the above analysis. We also list some results in the case of smaller size (see the last two rows in Fig. 7). The difference of micromagnetic configurations among the various DMI settings is more notable.

**Conclusion**

We have presented a comprehensive study of the DMI effect on the double-interface STT-MTJ. As is well known, the double-interface MTJ was developed for enhancing the thermal stability barrier. In this work our results prove another advantage of double-interface MTJ, that is, suppressing the detrimental effect of the DMI. The DMIs in two free layers could be suppressed or even cancelled out if they are configured with appropriate values and opposite signs, which is naturally satisfied by the double-interface STT-MTJ structure. A theoretical model was proposed to explain the conclusion. Micromagnetic results were discussed for revealing the roles of DMI played in the magnetization dynamics. Our work provides a feasible approach of minimizing the DMI in the double-interface STT-MTJ.

**List of abbreviations**

MTJ: Magnetic tunnel junction; DMI: Dzyaloshinskii-Moriya interaction; RKKY: Ruderman–Kittel–Kasuya–Yosida; STT: Spin transfer torque; MRAM: Magnetic random-access memory; FM/HM: ferromagnet/heavy-metal; SOC: spin-orbit coupling; FL: free layer.

**Availability of Data and Materials**

All data are fully available without restriction.

**Competing Interests**

The authors declare that they have no competing interests.

**Funding**

The work was partly supported by the National Natural Science Foundation of China under Grant 61704005.

**Authors' Contributions**

WZH and ZWS conceived and supervised the project. LSM and WYJ performed the simulation. LSM, WZH, and WMX wrote the paper and analyzed the results. All the authors have read and approved the final manuscript.

**Acknowledgement**

The authors would like to thank Xianjie Tang and Haoyang Zhang for their technical help.

**Table 1.** Parameters used in simulation

| Parameters | Description | Value |
|---|---|---|
| $M_s$ | Saturation magnetization | 1 MA/m |
| d | MTJ diameter | 40 nm |
| α | Gilbert damping constant | 0.01 |
| P | Spin polarization | 0.5 |
| J | Applied current density | 4 MA/cm$^2$ |
| A | exchange stiffness | 20 pJ/m$^2$ |
| $K_{u1}$ | Anisotropy constant of FL1 | 0.8 MJ/m$^3$ |
| $K_{u2}$ | Anisotropy constant of FL2 | 0.7 MJ/m$^3$ |
| $t_1$ | Thickness of FL1 | 1 nm |
| $t_2$ | Thickness of FL2 | 1.5 nm |
| $D_1$ and $D_2$ | DMI magnitudes of FL1 and FL2 | -2 to 2 mJ/m$^2$ |
| σ | Bilinear surface exchange energy for RKKY interaction | $3\times10^{-4}$ J/m$^2$ to $10^{-2}$ J/m$^2$ |

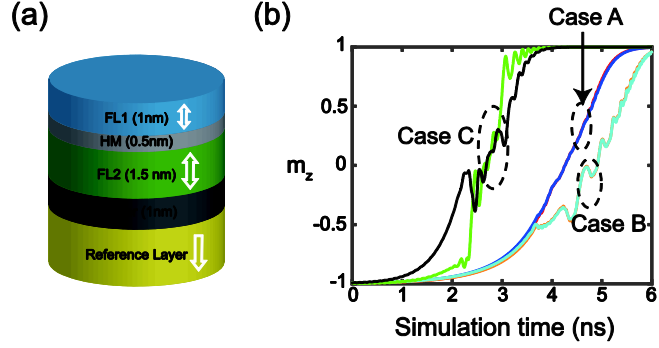

**Fig. 1.** (a) Schematic structure of the device studied in this work. The other layers are not shown for the clarity. (b) Typical results of the time-dependent $m_z$ (perpendicular-component of the unit magnetization). Case A: $\sigma = 1\times10^{-3}$ J/m$^2$, $D_1 = D_2 = 0$ (red for FL1, blue for FL2). Case B: $\sigma = 1\times10^{-3}$ J/m$^2$, $D_1 = 1$ mJ/m$^2$, $D_2 = -1$ mJ/m$^2$ (orange for FL1, cyan for FL2). Case C: $\sigma = 1\times10^{-4}$ J/m$^2$, $D_1 = D_2 = 0$ (green for FL1, black for FL2).

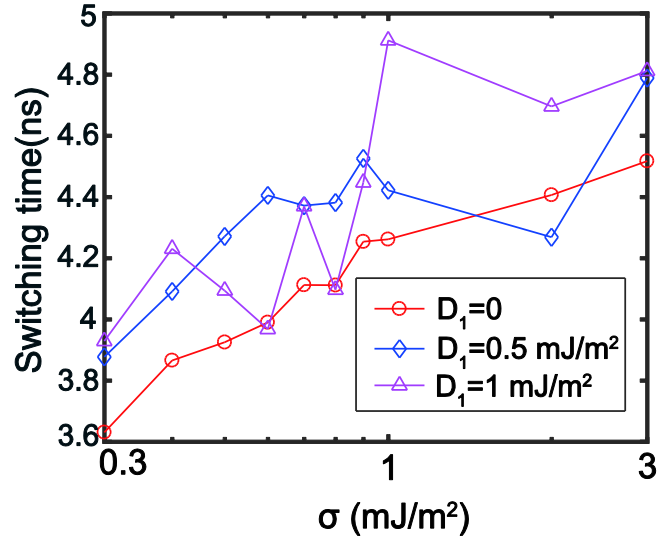

**Fig. 2.** Switching time as a function of RKKY strength, with $\sigma$ shown in the logarithm scale. $D_1$ and $D_2$ are set to the same values, but with the opposite signs.

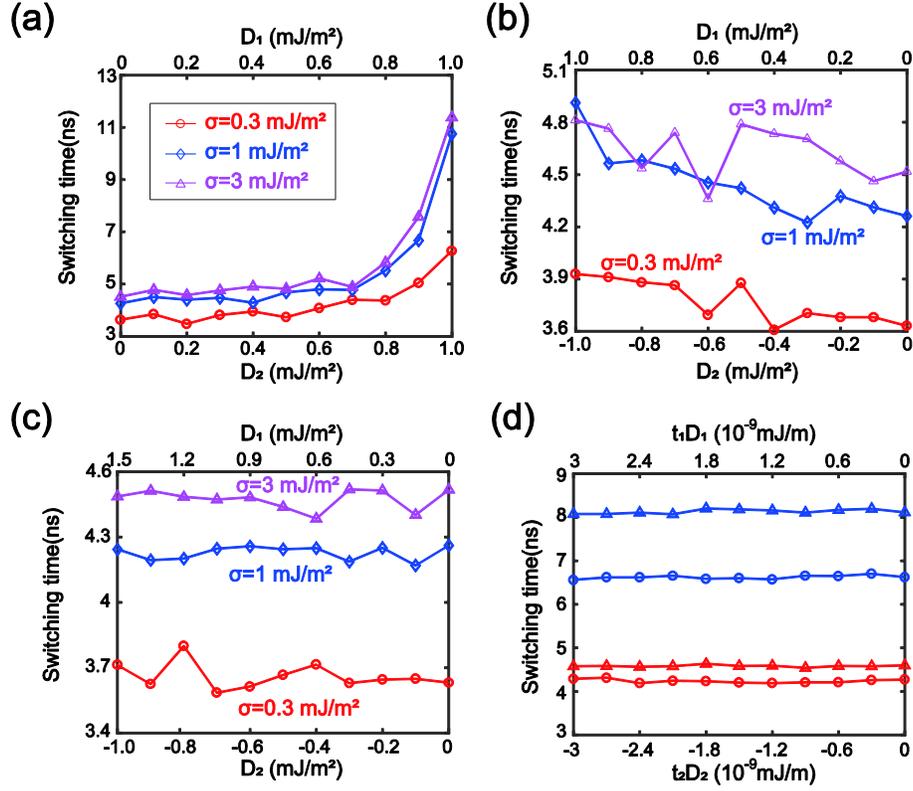

**Fig. 3.** Switching time as a function of DMI strength. (a) $D_1$ and $D_2$ are set to the same positive value. (b) $D_1$ and $D_2$ are set to the same value, but with the opposite signs. (c) $D_1$ and $D_2$ are configured to meet $t_1 D_1 + t_2 D_2 = 0$. (d) Additional results while changing the thickness or anisotropy constant, meanwhile keeping $t_1 D_1 + t_2 D_2 = 0$. blue line: $t_1$ is changed to 2 nm; red line: $t_1$ is changed to 1.5 nm. Triangle data: $\sigma = 3 \times 10^{-3}$ J/m². Circle data: $\sigma = 1 \times 10^{-3}$ J/m².

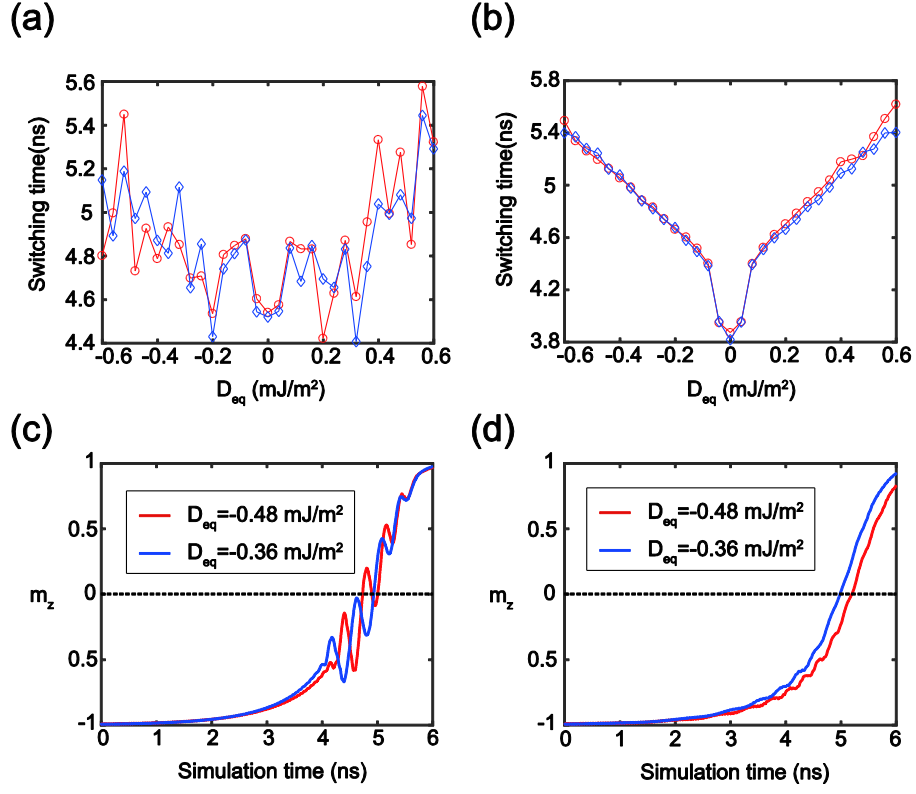

**Fig. 4.** (a)-(b) Switching time as a function of $D_{eq}$. Each $D_{eq}$ is obtained with two pairs of different $\{D_1, D_2\}$ values according to Eq. (3). Red curve: $D_1$ is varied meanwhile $D_2$ is fixed to 1 mJ/m². Blue curve: $D_1$ and $D_2$ are always set to the same value. Here $\sigma = 1 \times 10^{-2}$ J/m². In (a), the other parameters are configured as Table 1. In (b), $K_{u1} = K_{u2} = 0.7$ MJ/m³ for an ideal case. (c) and (d): Typical results of time-dependent $m_z$ corresponding to (a) and (b), respectively.

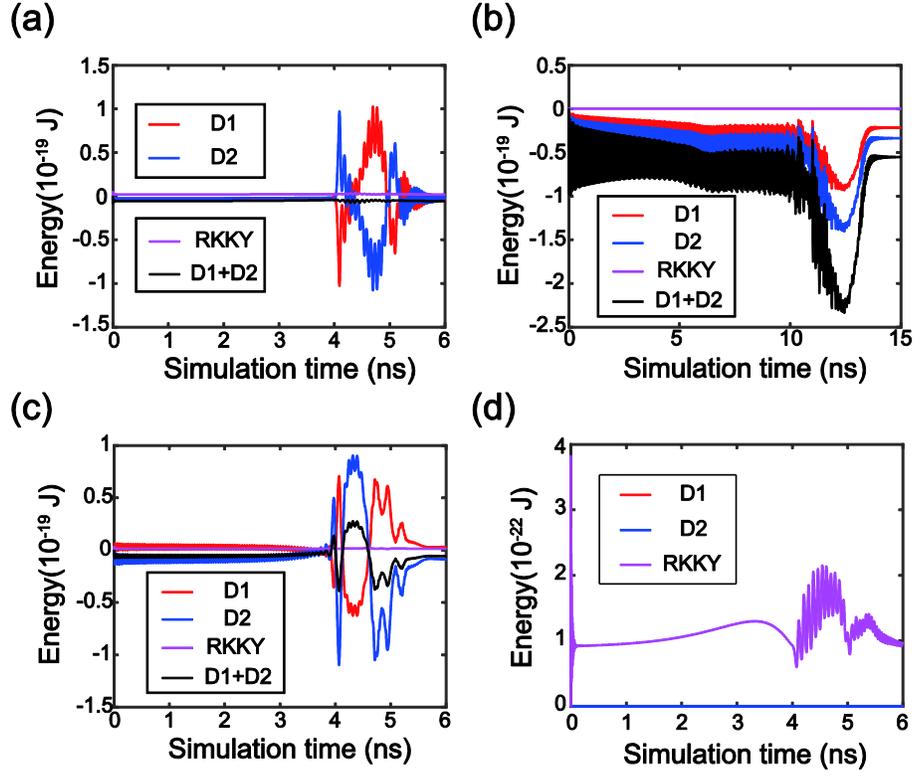

**Fig. 5.** Time evolution of the DMI and RKKY energies. (a) $D_1 = 1.5 \text{ mJ/m}^2$, $D_2 = -1 \text{ mJ/m}^2$, i.e. DMI effect is cancelled out; (b) $D_1 = D_2 = 1 \text{ mJ/m}^2$, i.e. DMI effect is accumulated; (c) $D_1 = 1 \text{ mJ/m}^2$, $D_2 = -1 \text{ mJ/m}^2$, i.e. DMI effect is mitigated but not cancelled out; (d) $D_1 = D_2 = 0$.

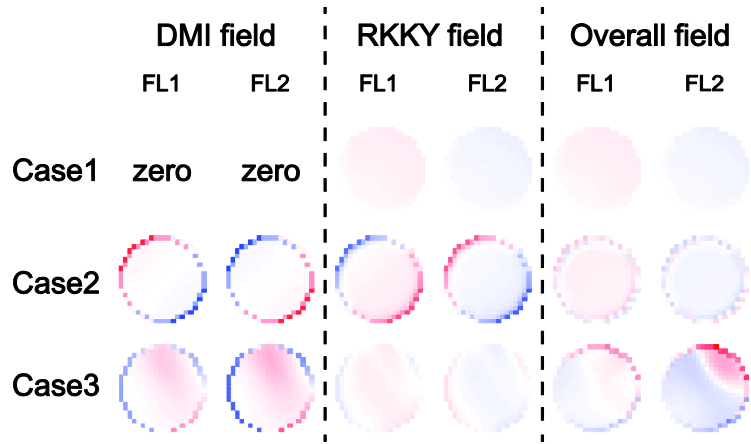

**Fig. 6.** Spatial distributions of the DMI and RKKY fields. Here a typical result at one time moment is shown for each case. The conclusion remains unchanged at the other time moments. Case 1: $D_1 = D_2 = 0$; Case 2: $D_1 = 1.5$ mJ/m$^2$, $D_2 = -1$ mJ/m$^2$, i.e. DMI effect is cancelled out; Case 3: $D_1 = D_2 = 1$ mJ/m$^2$, i.e. DMI effect is accumulated.

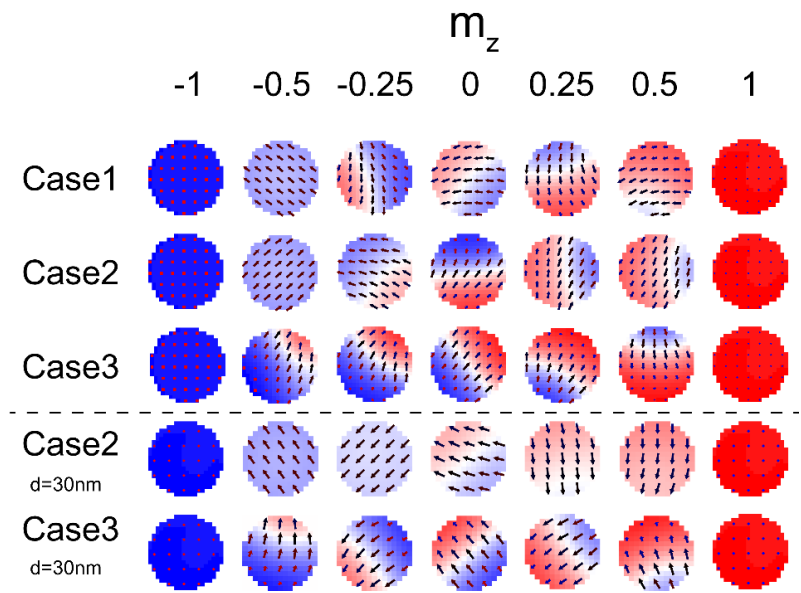

**Fig. 7.** Micromagnetic configurations during the magnetization switching. Here Cases 1~3 are configured the same parameters as Fig. 6.

19